\documentclass[pdflatex,sn-nature]{sn-jnl}
\usepackage{mathrsfs}

\usepackage{graphicx}%
\usepackage{multirow}%
\usepackage{amsmath,amssymb,amsfonts}%
\usepackage{amsthm}%
\usepackage{mathrsfs}%
\usepackage[title]{appendix}%
\usepackage{xcolor}%
\usepackage{textcomp}%
\usepackage{manyfoot}%
\usepackage{booktabs}%
\usepackage{algorithm}%
\usepackage{algorithmicx}%
\usepackage{algpseudocode}%
\usepackage{listings}%
\usepackage{array}
\usepackage{multirow}
\usepackage{makecell}
\usepackage{rotating}
\usepackage[table]{xcolor}
\usepackage{caption}
\usepackage{ragged2e}
\usepackage{afterpage}

\renewcommand{\arraystretch}{1.5}
\definecolor{lightgray}{gray}{0.9}

\theoremstyle{thmstyleone}%
\theoremstyle{thmstyletwo}%

\theoremstyle{thmstylethree}%

\begin{document}

\title[Article Title]{Scalable ultrafast random bit generation using wideband chaos-based entropy sources}


\author*[1,2]{\fnm{Chin-Hao} \sur{Tseng}}\email{chtseng0828@mail.saitama-u.ac.jp}

\author[1]{\fnm{Atsushi} \sur{Uchida}}\email{auchida@mail.saitama-u.ac.jp}

\author*[2,3]{\fnm{Sheng-Kwang} \sur{Hwang}}\email{skhwang@mail.ncku.edu.tw}


\affil[1]{\orgdiv{Department of Information and Computer Sciences}, \orgname{Saitama University}, \orgaddress{\street{\\255 Shimo-Okubo, Sakura-ku}, \city{Saitama City}, Saitama, \postcode{338-8570}, \country{Japan}}}

\affil[2]{\orgdiv{Department of Photonics}, \orgname{National Cheng Kung University}, \orgaddress{\street{\\1 University Road}, \city{Tainan City}, \postcode{701401}, \country{Taiwan}}}

\affil[3]{\orgdiv{Meta-nanoPhotonics Center}, \orgname{National Cheng Kung University}, \orgaddress{\street{\\1 University Road}, \city{Tainan City}, \postcode{701401}, \country{Taiwan}}}


\abstract{

The exponential growth of data transmission and processing speeds in modern digital infrastructure requires entropy sources capable of producing large volumes of true randomness for information security. Chaotic emissions from semiconductor lasers are attractive in this context because of their fast dynamics and nonrepetitive behavior. Their spectral bandwidth, however, is typically limited to several tens of gigahertz, which constrains the achievable entropy rate and makes ultrafast random bit generation difficult without substantial post-processing. Here, we demonstrate a chaos-based entropy source that employs optical heterodyning between the chaotic emission from a semiconductor laser and an optical frequency comb, yielding a bandwidth exceeding 100~GHz and an experimentally verified single-channel entropy rate of 1.86~Tb/s. By directly extracting multiple bits from the digitized output of the entropy source, we achieve a single-channel random bit generation rate of 1.536~Tb/s, while four-channel parallelization reaches 6.144~Tb/s with no observable interchannel correlation. This linear scalability suggests that aggregate throughput could reach hundreds of terabits per second with additional parallel channels. The broadband, low-overhead photonic architecture presented here provides a viable route to real-time, ultrafast random bit generation with broad implications for secure communications, high-performance AI computing, and large-scale data analytics.
}

\maketitle
\section*{Introduction}\label{sec1}

The rapid expansion of modern digital information infrastructure, driven by cloud computing, artificial intelligence (AI), and the Internet of Things (IoT), continues to push data transmission and processing toward ever higher speeds~\cite{zhang2020empowering}. To protect information against unauthorized access during transmission and processing, large volumes of trusted random numbers are required for cryptographic key generation~\cite{gisin2002quantum,yoshimura2012secure,koizumi2013information,gao20210}. Notably, data centers are expected to transmit and process information at rates exceeding a terabit per second in the near future~\cite{st2024practical}, which, in principle, calls for random number generation at comparable speeds to support secure line-rate encryption. Beyond cryptographic security, high-speed random numbers with certified randomness also support a broad range of applications, including generative adversarial networks~\cite{naruse2019generative}, large-scale AI computing, Monte Carlo simulations~\cite{metropolis1949monte}, and stochastic process modelling~\cite{asmussen2007stochastic}, where they enhance model creativity, output diversity, and computational accuracy. Meeting these rapidly growing and diverse demands calls for entropy sources that are fast, reliable, fundamentally unpredictable, and capable of delivering nondeterministic random numbers at high throughput.

Chaos, a universal phenomenon characterized by extreme sensitivity to initial conditions, provides complex and inherently non-repetitive signals~\cite{ott2002chaos,uchida2012optical}. These properties make it valuable for LiDAR~\cite{lin2004chaotic,chen2023breaking}, RADAR~\cite{lin2004radar,tseng2020broadband}, secure communication~\cite{argyris2005chaos}, photonic computing~\cite{naruse2017ultrafast,naruse2018scalable,iwami2022controlling}, and high-speed physical random bit generation (RBG)~\cite{uchida2008fast}. Chaotic semiconductor lasers have emerged as promising entropy-source candidates because their temporal evolution is quantum-noise-dependent~\cite{ornstein1989ergodic,mikami2012estimation,harayama2012theory,sunada2012noise}, they exhibit rapid relaxation oscillations~\cite{vahala1983observation}, and they are compatible with photonic integration. The first demonstration of chaotic-laser-based RBG achieved a generation rate of 1.7~Gb/s using two independent lasers~\cite{uchida2008fast}. Subsequent works~\cite{reidler2009ultrahigh,kanter2010optical,hirano2010fast,argyris2010implementation,akizawa2012fast,Li:12,li2013heterodyne,yamazaki2013performance,li2014two,sakuraba2015tb,tang2015tbits,rontani2016enhanced,wang2017minimal,ugajin2017real,li2017real,zhang2017640,tseng2021high,ruan2023simultaneous} have increased generation rates to the hundreds of Gb/s regime by broadening the chaos bandwidth and incorporating digital post-processing. Although the chaos bandwidth of semiconductor lasers has reached approximately 40~GHz~\cite{zhang2017640,qiao2019generation,yang2020generation}, surpassing many electronic~\cite{wang2016theory,gong2019true} and photonic approaches, the corresponding entropy throughput~\cite{hart2017recommendations,yoshiya2020entropy,kawaguchi2021entropy,tseng2021high,tseng2024entropy} remains insufficient for terabit-per-second RBG. Several studies have reported Tb/s RBG using complex offline post-processing~\cite{li2014two,sakuraba2015tb,tang2015tbits,butler2016optical,ruan2023simultaneous,huang2024parallel,zhang2024multi} to compensate for limited entropy at the sources. Such methods introduce latency and hardware overhead, which may restrict real-time operation. Entropy sources with flat spectra extending beyond 100~GHz are needed to directly support terabit-per-second entropy rates, since the achievable entropy rate is fundamentally tied to both available bandwidth and spectral flatness~\cite{hart2017recommendations,tseng2024entropy}. Achieving flat and broad spectra remains challenging because the bandwidth of chaotic semiconductor lasers is constrained by their relaxation oscillation frequency, typically near 10~GHz.

Parallel RBG architectures provide another means to reach high aggregate rates by operating multiple uncorrelated channels simultaneously. Several methods have demonstrated terabit-per-second throughput using a few to hundreds of channels, including interference of spatiotemporal modes~\cite{kim2021massively}, chaotic microcombs~\cite{shen2023harnessing,hu2023massive}, and chaotic semiconductor lasers~\cite{sakuraba2015tb,tang2015tbits,xiang20192,han2020generation,cai2023tbps,ruan2023simultaneous,huang2024parallel,zhang2024multi}. Many of these approaches exhibit measurable interchannel correlations, which require complex post-processing to maintain statistical independence, increasing system complexity and limiting real-time operation. Systems employing independent lasers~\cite{huang2024parallel} can mitigate these correlations, however, their entropy-source bandwidth per channel is typically below 30~GHz, which fundamentally restricts overall entropy throughput. This underscores the need for entropy sources with intrinsically high entropy rates and scalability across parallel channels.

This paper presents an experimentally validated wideband chaos-based entropy source (WCBES) that achieves a spectral bandwidth exceeding 100~GHz through optical heterodyning between the chaotic emission of a semiconductor laser and an optical frequency comb. To the best of our knowledge, this is the highest experimentally reported bandwidth and entropy rate (1.86~Tb/s) for any chaos-based entropy source. Direct multi-bit extraction from the digitized WCBES output enables a single-channel RBG rate of 1.536~Tb/s. Using the broadband and naturally uncorrelated characteristics of each WCBES channel, we further demonstrate a four-channel parallel RBG system with an aggregate rate of 6.144~Tb/s. With additional channel parallelization, this approach could reach aggregate throughput in the hundreds of terabits per second. The proposed photonic architecture provides a promising route toward real-time ultrafast RBG for secure communications, high-performance AI computing, large-scale stochastic simulations, and emerging information technologies.

\section*{Results}\label{sec2}
\subsection*{Wideband chaos-based entropy source generation}

We experimentally demonstrate a WCBES based on optical heterodyning between the chaotic emission from a semiconductor laser and an optical frequency comb generated by a comb laser. The experimental setup is shown in Fig.~\ref{fig1}a, with detailed descriptions provided in Supplementary Note~I. Before discussing the heterodyne process, we first examine the chaotic dynamics of the laser diode subject to optical feedback. Operating in a chaotic regime under delayed optical self-feedback, this laser diode produces fast and irregular intensity oscillations (Fig.~\ref{fig1}b, upper blue trace), arising from rapid relaxation oscillations~\cite{vahala1983observation} and the nonlinear amplification and mixing of microscopic noise~\cite{ornstein1989ergodic,mikami2012estimation,harayama2012theory,sunada2012noise}. By optimizing the feedback strength and bias current (see Supplementary Note~II), the electrical spectrum of the resulting chaotic laser (Fig.~\ref{fig1}c, upper blue trace) reaches a standard bandwidth~\cite{lin2012effective} of 40~GHz, an effective bandwidth~\cite{lin2012effective} of 38~GHz, and a spectral flatness~\cite{johnston2002transform} of 0.96. The latter value approaches the ideal flatness of 1 and indicates a highly uniform spectral distribution. These characteristics exceed those reported in earlier demonstrations of optical-feedback-induced chaos in semiconductor lasers, which typically achieve only a few tens of gigahertz and exhibit nonuniform spectral profiles, thereby limiting the achievable entropy throughput. Although the chaotic laser already exhibits high bandwidth and flatness (Fig.~\ref{fig1}c, upper blue trace), generating entropy at terabit-per-second throughput for ultrafast RBG requires a substantially broader spectrum with comparably high flatness, as suggested by the Shannon–Hartley theorem~\cite{hart2017recommendations,tseng2024entropy}. To meet this requirement, we employ optical heterodyning between the chaotic laser output and an optical frequency comb

\clearpage

\begin{figure}[t]
\centering
\includegraphics[width=0.85\textwidth]{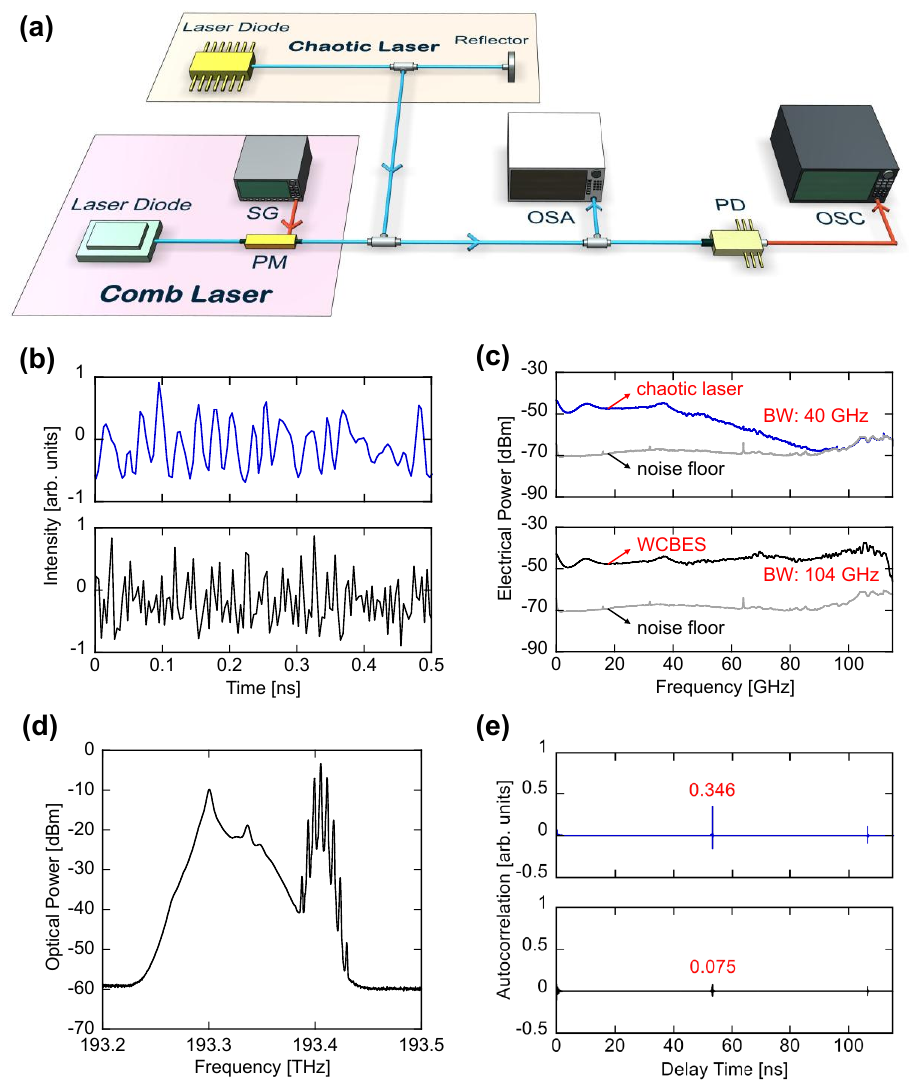}
\caption{
\textbf{Generation of the wideband chaos-based entropy source (WCBES).}
\textbf{a}, Schematic of the experimental setup. OSA, optical spectrum analyzer; OSC, real-time digital oscilloscope; PM, phase modulator; PD, photodetector; SG, signal generator.
\textbf{b}, Intensity fluctuations of the emission from the chaotic laser (upper blue trace) and the WCBES (lower black trace), showing substantially faster temporal oscillations in the WCBES.
\textbf{c}, Electrical spectra of the emission from the chaotic laser (upper blue trace) and the WCBES (lower black trace). The chaotic laser exhibits a standard bandwidth of 40~GHz, an effective bandwidth of 38~GHz, and a spectral flatness of 0.96, whereas the WCBES has a standard bandwidth of 104~GHz, an effective bandwidth of 73~GHz, and a spectral flatness of 0.91.
\textbf{d}, Optical spectrum of the combined output from the chaotic laser and the comb laser.
\textbf{e}, Autocorrelation functions of the emission from the chaotic laser (upper blue trace) and the WCBES (lower black trace), showing a reduction in the time-delay signature from 0.346 to 0.075, indicating enhanced temporal randomness in the WCBES.
}
\label{fig1}
\end{figure}

\clearpage

\noindent generated by a comb laser (realized via phase modulation of another semiconductor laser diode) to further expand the spectrum while preserving uniformity.

Figure~\ref{fig1}d presents the optical spectrum of the combined signal formed by mixing the chaotic emission with the optical frequency comb. Heterodyne detection of this combined optical field generates a WCBES with much faster temporal oscillations (Fig.~\ref{fig1}b, lower black trace) than those of the chaotic laser alone. The resulting electrical spectrum (Fig.~\ref{fig1}c, lower black trace) exhibits a standard bandwidth of 104~GHz, an effective bandwidth of 73~GHz, and a spectral flatness of 0.91. To our knowledge, this is the broadest experimentally demonstrated chaos-based entropy source, achieved entirely with commercially available components. Broadband spectral expansion arises from heterodyne photomixing, where interference between the chaotic field and multiple comb lines converts fine chaotic phase variations into ultrafast intensity fluctuations. Numerical simulations further show that the heterodyne process introduces additional high-frequency components, thereby enhancing the temporal oscillation rate (see Supplementary Note~III). In addition, because the optical comb is generated via phase modulation, beating among the comb lines does not produce discrete spectral peaks in the electrical domain; this behavior is also confirmed by simulations (see Supplementary Note~III). This property makes the proposed approach well suited for extending chaos bandwidths while preserving intrinsic randomness. The measured bandwidth is ultimately limited by the photodetector and oscilloscope, whereas simulations indicate that the intrinsic spectrum can extend beyond 150~GHz (see Supplementary Note~III for a description of the spectral broadening mechanism and numerical results).

Beyond spectral broadening, the temporal randomness of the WCBES is further enhanced by the suppression of deterministic features in the chaotic waveform. Autocorrelation analysis reveals a substantial reduction in the time-delay signature (TDS)~\cite{hirano2010fast,zhang2017640,han2020generation}, defined as the dominant side peak in the autocorrelation function, from 0.346 in the self-feedback configuration (Fig.~\ref{fig1}e, upper blue trace) to 0.075 in the WCBES configuration (lower black trace). Previous work~\cite{zhang2017640} has demonstrated that RBG can be achieved using the widely adopted least-significant-bit (LSB) extraction method when TDS is low. Our WCBES achieves a sufficiently small TDS while simultaneously providing a flat spectrum exceeding 100~GHz, highlighting its robustness as an entropy source for high-throughput RBG.

\subsection*{Entropy rate assessment and random bit generation}

\begin{figure}[t]
\centering
\includegraphics[width=1.0\textwidth]{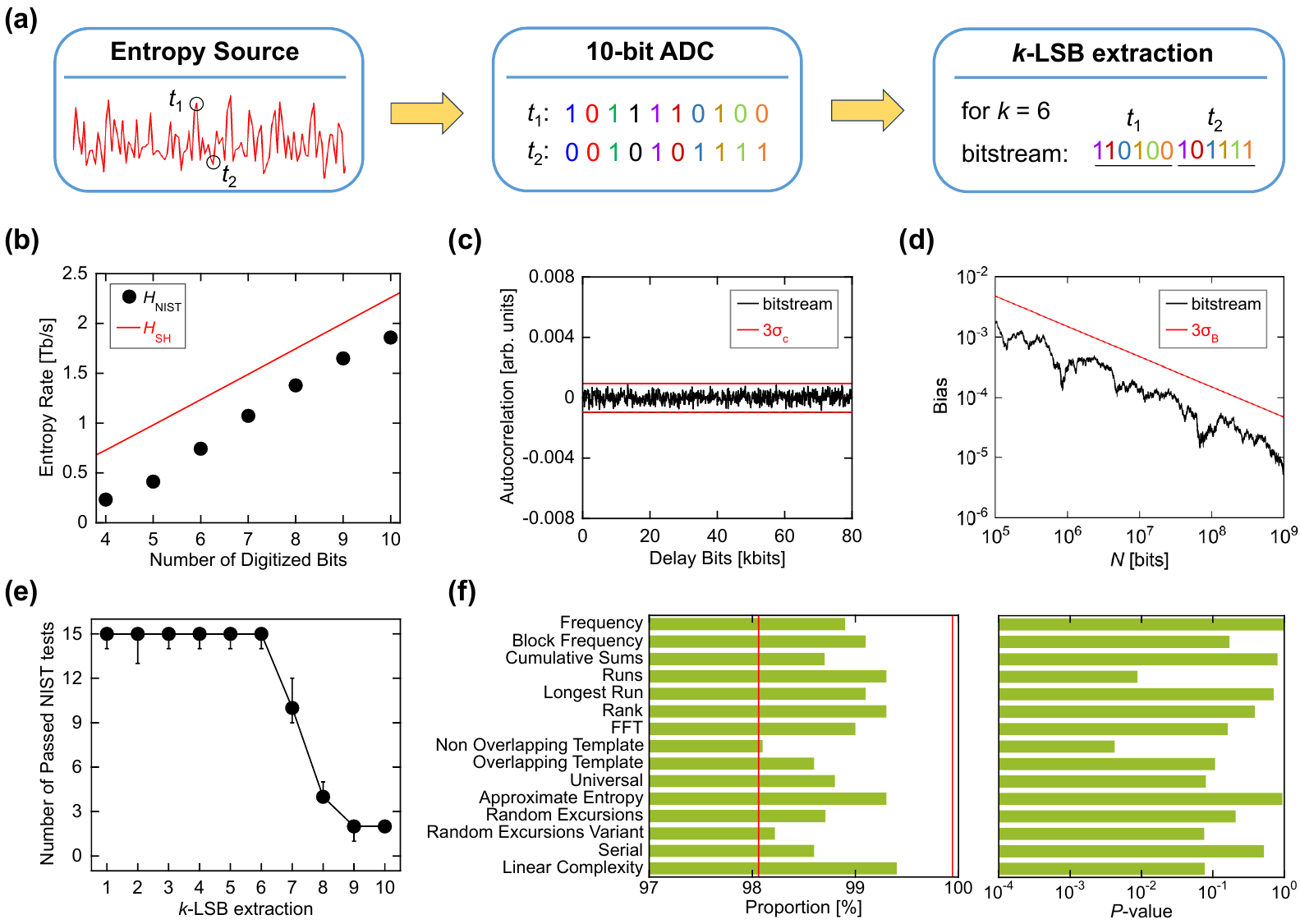}
\caption{
\textbf{Entropy rate assessment and random bit generation.}
\textbf{a}, Schematic of the data acquisition and bit-extraction pipeline. ADC, analog-to-digital converter; LSB, least significant bit.
\textbf{b}, Estimated entropy rate $H_{\rm NIST}$ as a function of the number of digitized bits $n$ (black symbols), evaluated using the NIST SP~800-90B entropy estimation suite. The red line denotes the Shannon–Hartley theoretical limit $H_{\rm SH}$.
\textbf{c}, Autocorrelation of the 6-LSB bitstream (black) with statistical bounds of $\pm3\sigma_{\rm c}$ (red), where $\sigma_{\rm c}=1/\sqrt{N}$ and $N=10^7$ bits.
\textbf{d}, Bias $|P_{1} - 0.5|$ as a function of $N$ (black), where $P_{1}$ is the probability of observing “1” in the sequence. The bias remains below the statistical threshold $3\sigma_{\rm B}$ (red), where $\sigma_{\rm B}=1/(2\sqrt{N})$, demonstrating long-term binary balance.
\textbf{e}, Number of passed NIST SP~800-22 tests as a function of the number of extracted LSBs $k$. A value of “15” indicates that all NIST tests are passed. For each $k$, eleven 1-Gbit sequences are evaluated; the median number of passed tests is plotted, with error bars showing the minimum and maximum values. The results indicate that LSBs with $k\le6$ can be reliably extracted from each digitized data sample for random bit generation.
\textbf{f}, Representative NIST SP~800-22 results for $k=6$, taken from one of the passed trials in \textbf{e}. All 15 pass proportions fall within the expected interval $0.99 \pm 0.0094$ (red) at a significance level of 0.01, and all $P$-values exceed 0.0001, confirming the statistical randomness of the 6-LSB bitstream.
}
\label{fig2}
\end{figure}

To evaluate the performance of the proposed WCBES for true RBG, we implement a complete data-acquisition and bit-extraction pipeline, as shown in Fig.~\ref{fig2}a. The WCBES output is sampled using a high-speed 10-bit analog-to-digital converter (ADC) operating at 256~GS/s. We first measure the entropy rate of the WCBES to assess its suitability for ultrafast RBG. Figure~\ref{fig2}b shows the estimated entropy rate $H_{\rm NIST}$ as a function of the number of digitized bits $n$, evaluated using the NIST SP~800-90B entropy estimation suite~\cite{NIST.SP.800-90B-2018}. As shown, $H_{\rm NIST}$ exceeds 1~Tb/s for $n \ge 7$, reaching 1.86~Tb/s at $n=10$. To the best of our knowledge, this is the highest experimentally

\clearpage

\clearpage

\noindent reported entropy rate from a chaos-based entropy source to date. This performance results from the broadband spectral coverage and high spectral flatness of the WCBES, where the former supports high-Nyquist-rate sampling and the latter increases unpredictability between samples, enabling ultrahigh entropy throughput. Notably, $H_{\rm NIST}$ remains below the Shannon–Hartley limit $H_{\rm SH}$ (Fig.~\ref{fig2}b, red trace)~\cite{hart2017recommendations}, confirming the validity of the estimation, and gradually approaches $H_{\rm SH}$ for $n \ge 7$, indicating near-optimal utilization of the source’s entropy capacity.

For RBG, $k$ LSBs are extracted directly from each digitized sample and concatenated to form a binary sequence (Fig.~\ref{fig2}a). This widely adopted multibit extraction method enables real-time operation with a generation rate that scales linearly with $k$. Here, we set $k=6$, yielding an RBG rate of 1.536~Tb/s ($=$ 256~GS/s $\times$ 6 LSBs) while maintaining randomness. Figure~\ref{fig2}c shows the autocorrelation of the resulting 6-LSB bitstream (black trace). No significant peaks appear beyond zero lag, and all correlation coefficients fall within the ±$3\sigma_{\rm c}$ bounds (red lines), where $\sigma_{\rm c}=1/\sqrt{N}$~\cite{li2014two,tang2015tbits,huang2024parallel} and $N=10^7$ bits. These results confirm that TDSs are effectively suppressed by LSB extraction~\cite{zhang2017640}, verifying the temporal independence of the generated bits.
The binary uniformity of the 6-LSB bitstream is examined by evaluating the bias $|P_{1} - 0.5|$ as a function of $N$ bits (Fig.~\ref{fig2}d), where $P_{1}$ denotes the probability of ‘1’ in the sequence. The bias remains well below the confidence bound $3\sigma_{\rm B}$ (red line), where $\sigma_{\rm B}=1/(2\sqrt{N})$~\cite{kanter2010optical,ugajin2017real}, confirming stable long-term binary balance.

Randomness is further assessed using the NIST SP~800-22 statistical test suite~\cite{NIST.SP.800-22rev1a-2010}, which includes 15 tests that must all be passed to demonstrate statistical randomness. Figure~\ref{fig2}e shows the number of passed tests as a function of the extracted $k$ LSBs. For each $k$, eleven independent 1-Gbit sequences are evaluated, with the median number of passed tests plotted and error bars indicating the full range. All 15 tests are passed for $k \le 6$, confirming that six LSBs can be reliably extracted from each digitized data sample.

A representative passing result for $k=6$ is presented in Fig.~\ref{fig2}f, corresponding to one of the validated trials in Fig.~\ref{fig2}e. At a significance level of 0.01, all 15 pass proportions lie within the expected interval $0.99 \pm 0.0094$ (red lines), and all $P$-values exceed 0.0001. These results confirm that the 1.536~Tb/s random bitstream generated by the WCBES passes all NIST SP~800-22 tests. The RBG rate also remains below the measured source entropy rate of 1.86~Tb/s ($H_{\rm NIST}$ at $n=10$), ensuring operation within the available entropy budget.

\clearpage

\subsection*{Parallel random-bit generation and multi-channel validation}

To further increase the entropy throughput, we implement a parallel architecture comprising $M$ independent WCBES units (Fig.~\ref{fig3}a), with the details of a single unit provided in Supplementary Note~I. The output from the comb laser is divided into $M$ optical paths, each combined with the chaotic emission from an independent semiconductor laser with optical self-feedback (Laser Diode~1 to $M$). Photodetection in each channel using $\mathrm{PD}_m$ ($m = 1,\dots,M$) produces electrically independent WCBESs. Nonlinear amplification of independent spontaneous emission noise seeds in each semiconductor laser ensures temporal independence among the channels. Each WCBES exhibits a standard bandwidth exceeding 100~GHz (black symbols), an effective bandwidth surpassing 72~GHz (green symbols), and a spectral flatness of approximately 0.9 (red symbols) in Fig.~\ref{fig3}b, consistent with the single-channel performance shown in Fig.~\ref{fig1}c. This intrinsic temporal independence highlights the scalability of the architecture. The present implementation is limited to $M=4$ by hardware constraints rather than by any fundamental restriction.

For entropy-rate analysis, each WCBES is digitized using a 10-bit ADC operating at 256~GS/s. At $n = 10$, $H_{\mathrm{NIST}}$ reaches approximately 1.86~Tb/s on all channels (Fig.~\ref{fig3}c), in agreement with the single-channel results in Fig.~\ref{fig2}b and confirming the scalability of the approach for multi-channel operation. For RBG, six LSBs are extracted from each channel to generate four independent random bitstreams. Statistical independence among channels is verified by cross-correlation analysis of the four bitstreams (Fig.~\ref{fig3}d). All off-diagonal correlation coefficients are within the statistical bounds of $\pm 3\sigma_{\mathrm{c}}$, with $N=10^7$ bits used in the estimation, indicating negligible interchannel correlation.

We further examine the binary uniformity of the four bitstreams. The maximum bias across channels (Fig.~\ref{fig3}e, black trace) remains well below the statistical threshold of $3\sigma_{\mathrm{B}}$ (red line) for all values of $N$, confirming the long-term binary balance required for parallel RBG. Each channel is then evaluated using the NIST SP~800-22 suite, following the procedure used for the single-channel case in Fig.~\ref{fig2}e. For each channel, the performance indicators are defined as the median of the minimum pass proportions and the median of the lowest $P$-values across eleven independent NIST SP~800-22 trials, where the minimum pass proportion and lowest $P$-value are taken from the 15 individual tests in each trial. The resulting minimum pass proportions (Fig.~\ref{fig3}f) fall within the acceptance interval $0.99 \pm 0.0094$ (red line), and the corresponding lowest $P$-values (Fig.~\ref{fig3}g) exceed 0.0001. These results confirm that the four parallel bitstreams satisfy the required statistical randomness criteria, yielding an aggregate RBG rate of 6.144~Tb/s ($=$ 4 channels~$\times$~1.536~Tb/s).

\begin{figure}[t]
\centering
\includegraphics[width=0.95\textwidth]{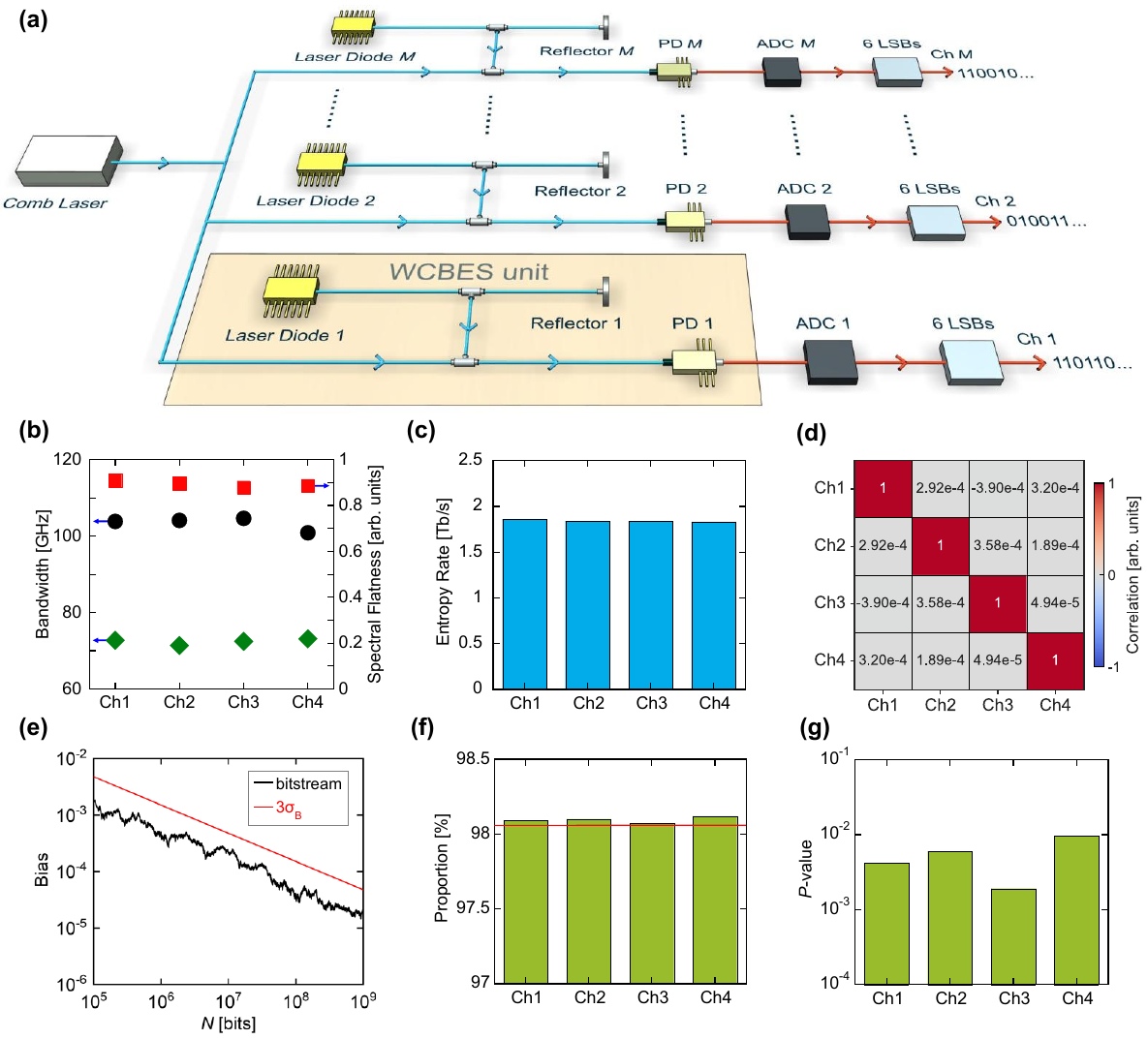}
\caption{
\textbf{Parallel RBG using multiple independent WCBES channels.}
\textbf{a}, Schematic of the scalable parallel RBG architecture. The output from a comb laser is split into $M$ optical paths, each combined with chaotic emission from an independent semiconductor laser. This generates $M$ physically distinct WCBESs, each detected by a high-speed PD and digitized by a 10-bit ADC at 256~GS/s. The present demonstration is limited to $M=4$ by currently available hardware.
\textbf{b}, Electrical spectral analysis of the four WCBES channels, each exhibiting a standard bandwidth exceeding 100~GHz (black symbols), an effective bandwidth over 72~GHz (green symbols), and a spectral flatness of approximately 0.9 (red symbols).
\textbf{c}, $H_{\mathrm{NIST}}$ for Channel 1 to 4 under 10-bit digitization. All values exceed 1.8~Tb/s, demonstrating consistent entropy generation across the four channels.
\textbf{d}, Cross-correlation coefficients between all channel pairs. Off-diagonal values lie within statistical bounds of $\pm3\sigma_{\mathrm{c}}$ with $N=10^{7}$~bits, indicating negligible inter-channel correlation.
\textbf{e}, Maximum bias across the four channels as a function of $N$ bits (black). Bias remains below the statistical threshold $3\sigma_{\mathrm{B}}$ (red), confirming long-term binary balance of the bitstreams from all channels.
\textbf{f}, Median value of the minimum pass proportion across eleven NIST SP~800-22 trials for each channel, all within the statistical acceptance range of $0.99 \pm 0.0094$ (red).
\textbf{g}, Median value of the lowest $P$-values across eleven NIST SP~800-22 trials for each channel, all exceeding 0.0001. Together, the results in \textbf{f} and \textbf{g} confirm that the bitstreams from all four channels satisfy the statistical randomness criteria.
}
\label{fig3}
\end{figure}

\clearpage

By bypassing current hardware limitations, the aggregate throughput can be extended to hundreds of terabits per second through additional parallel channels. Recent advances in photonic integration, particularly the miniaturization of self-feedback semiconductor lasers~\cite{argyris2010implementation,ugajin2017real} and the development of laser arrays~\cite{ma2024photonic,wang2024optical}, provide promising routes for large-scale integration of multiple WCBES units. Such architectures can support compact RBG systems operating at hundreds of terabits per second. These results demonstrate both the scalability and practical potential of the proposed parallel approach, enabling real-time and massively parallel true random number generation for cryptographic and high-performance computing applications.

\vspace{1em}

\section*{Discussion}

We benchmark our WCBES against established photonic RBG architectures. Key performance indicators, including spectral bandwidth, entropy rate, RBG throughput, number of demonstrated channels, and reliance on post-processing are summarized in Table~\ref{tab:RBG_comparison}. This comparison highlights the distinguishing features and advantages of the proposed approach, particularly its broader spectral bandwidth, higher entropy throughput, and the ability to generate certified random bits through direct multi-bit extraction from the digitized WCBES output.

Both single-channel capability and multichannel scalability are critical performance factors. First, we examine single-channel characteristics. For chaos-based RBG using semiconductor lasers, various perturbation schemes have been developed to broaden the chaos bandwidth, thereby enhancing the entropy rate. Reported single-channel bandwidths reach up to 40~GHz~\cite{zhang2017640,qiao2019generation,yang2020generation}, with entropy rates approaching 500~Gb/s~\cite{tang2015tbits}. Certain schemes~\cite{li2014two,huang2024parallel} have demonstrated RBG rates exceeding 2~Tb/s per channel, however, such rates surpass the intrinsic entropy of the chaotic source and rely on complex post-processing (e.g., high-order derivative operations) to compensate for insufficient entropy generation. These operations introduce additional hardware overhead and latency, which can limit real-time implementation.

Entropy sources based on chaotic microcombs~\cite{shen2023harnessing,hu2023massive} or quantum emitters~\cite{bruynsteen2023100} are generally constrained by narrow spectral bandwidths, limiting both entropy throughput and achievable RBG rates per channel. Recently, amplified spontaneous emission (ASE) from laser diodes~\cite{kim2021massively} has shown broadband spectra exceeding 300~GHz and theoretically reported single-channel RBG rates of around 2~Tb/s. In this demonstration, the temporal waveform was reconstructed from streak-camera images (each image covering only 0.5~ns in duration), spanning multiple frames, and random bits were generated through image-based post-processing involving pixel binning, delay differences, and exclusive-OR (XOR) operations. This reliance on multi-frame reconstruction and computational post-processing can restrict real-time acquisition and integration into practical photonic systems.

For multichannel RBG, several approaches have demonstrated parallel implementations with terabit-per-second throughput. Chaotic semiconductor lasers with five parallel channels achieved 4.64~Tb/s~\cite{zhang2024multi}, while chaotic microcombs with 32 channels demonstrated 3.84~Tb/s~\cite{shen2023harnessing} and could, in principle, scale to over 800 channels~\cite{hu2023massive}. However, correlations among channel pairs in these approaches are often non-negligible, necessitating complex post-processing to decorrelate the outputs and ensure channel independence. Similarly, ASE-based systems theoretically achieve 127 channels, however, these channels are defined through image segmentation and post-processing rather than independent physical channels, which may limit their practicality for scalable photonic implementation.

In this study, we experimentally demonstrate a WCBES with an intrinsic standard bandwidth exceeding 104~GHz, representing the broadest reported bandwidth for a chaos-based entropy source to date. The system achieves an experimentally validated entropy rate of 1.86~Tb/s and a single-channel RBG rate of 1.536~Tb/s by directly extracting multiple LSBs from the digitized WCBES output. All entropy-source data are obtained through direct digital acquisition rather than offline reconstruction, ensuring faithful capture of the underlying physical chaos. Furthermore, our parallel WCBES architecture comprises true physical channels that are intrinsically independent, each achieving an entropy rate of approximately 1.86~Tb/s, surpassing all previously reported chaos-based approaches. Although the current demonstration is limited to four channels due to hardware constraints, the system achieves an aggregate RBG rate of 6.144~Tb/s. The architecture is inherently scalable and, with further parallelization, could support over 200 channels (see Supplementary Section III), with projected aggregate entropy rates exceeding 372~Tb/s.
These results demonstrate the potential of WCBES to improve the performance of photonic RBG. The architecture provides a practical and scalable framework that reduces system complexity and latency, offering a pathway for future developments in secure communications, photonic AI computing, and large-scale data analytics.


\renewcommand\cellgape{\Gape[2pt]}  

\begin{sidewaystable}
\captionsetup{justification=raggedright, singlelinecheck=false}
\caption{Comparison of various photonic random bit generation (RBG) schemes.}
\label{tab:RBG_comparison}

\renewcommand{\arraystretch}{1.35}
\arrayrulecolor{black}

\begin{tabular*}{\textheight}{@{\extracolsep\fill}lllllll}
\toprule
\makecell[l]{\textbf{Scheme}} &
\makecell[l]{\textbf{Bandwidth}\\\textbf{(GHz)}} &
\makecell[l]{\textbf{Entropy Rate}\\\textbf{per Channel}\\\textbf{(Tb/s)}} &
\makecell[l]{\textbf{RBG Rate}\\\textbf{per Channel} \\\textbf{(Tb/s)}} &
\makecell[l]{\textbf{Number}\\\textbf{of Channel}} &
\makecell[l]{\textbf{Total RBG}\\\textbf{Rate (Tb/s)}} &
\makecell[l]{\textbf{Complex}\\\textbf{Post-Processing}\textsuperscript{\textbf{e,f}}} \\
\midrule
\arrayrulecolor{gray!50}

\makecell[l]{{Chaotic Laser}~\cite{li2014two}\\(self-feedback)} 
    & $<10.0^{\text{a}}$ & 0.176 & 2.2 & 1 & 2.2 & \makecell[l]{Yes\\(62nd-order derivatives)}\\

\hline
\makecell[l]{{Chaotic Laser}~\cite{sakuraba2015tb}\\(cascade injection)} 
    & 35.2\textsuperscript{a} & -- & 0.6 & 2 & 1.2 & \makecell[l]{Yes\\(bit-order reversal, XOR)} \\

\hline
\makecell[l]{{Chaotic Laser}~\cite{tang2015tbits}\\(mutual injection)} 
    & $<11.0^{\text{a}}$ & 0.495 & 0.56 & 2 & 1.12 & \makecell[l]{Yes\\(bit-order reversal, XOR)} \\

\hline
\makecell[l]{{Chaotic Laser}~\cite{xiang20192}\\(laser network)} 
    & $<20.0^{\text{a}}$ & -- & 0.32 & 7 & 2.24 & No \\

\hline
\makecell[l]{{Chaotic Laser}~\cite{huang2024parallel,tseng2024entropy}\\(modulated injection)} 
    & 30.0\textsuperscript{a} & 0.386 & 3.76 & 4 & 15.04 & \makecell[l]{Yes\\(52nd-order derivatives)} \\

\hline
\makecell[l]{{Chaotic Laser}~\cite{zhang2024multi}\\(phase modulation \&\\CFBG feedback)} 
    & 22.6--31.7\textsuperscript{a} & -- & \makecell[l]{0.8 (1 CH)\\0.96 (4 CH)} & 5 & 4.64 & \makecell[l]{Yes\\(delay difference, XOR)} \\

\hline
\makecell[l]{{Chaotic Combs}~\cite{shen2023harnessing}} 
    & 9.6\textsuperscript{b} & -- & 0.12 & 32 & 3.84 & \makecell[l]{Yes\\(delay difference, XOR)} \\

\hline
\makecell[l]{{Chaotic Combs}~\cite{hu2023massive}} 
    & $<2.5^{\text{a}}$ & -- & 0.012 & 6 $\text{(805}^\text{g}$) & 0.072 $\text{(9.66}^\text{g}$) & \makecell[l]{Yes\\(4th-order derivatives, XOR)} \\

\hline
\makecell[l]{{Quantum Emitter}~\cite{bruynsteen2023100}} 
    & $<8.0^{\text{a}}$ & 0.102 & 0.1 & 1 & 0.1 & No \\

\hline
\makecell[l]{{ASE of EDFA}~\cite{argyris2012sub}} 
    & $<12.0^{\text{a}}$ & -- & 0.56 & 1 & 0.56 & No \\

\hline
\makecell[l]{{ASE of Laser Diode}~\cite{kim2021massively}} 
    & \makecell[l]{315\textsuperscript{a}\\$\text{(concatenation}^\text{c}$)} & -- & 2 & \makecell[l]{127\\$\text{(Image-defined}^\text{d}$)} & 254 & \makecell[l]{Yes\\(image reconstruction,\\~delay difference, XOR)} \\

\arrayrulecolor{black}
\hline
\makecell[l]{\textbf{This Work}} 
    & \textbf{104\textsuperscript{a}} & \textbf{1.86} & \textbf{1.536} & \textbf{4} & \textbf{6.144} & \textbf{No} \\

\arrayrulecolor{black}
\botrule
\end{tabular*}

\vspace{2mm}
\footnotetext{\textsuperscript{a}\,Standard bandwidth, defined as the frequency range from DC containing 80\% of the total power.}
\footnotetext{\textsuperscript{b}\,10-dB bandwidth, defined as the frequency range from DC to the frequency at which the power drops by 10~dB from the peak value.}
\footnotetext{\textsuperscript{c}\,“Concatenation” indicates that the spectrum was estimated from discrete time windows rather than from a continuously sampled waveform.}
\footnotetext{\textsuperscript{d}\,“Image-defined” denotes channels obtained by image processing rather than by physical separation.}
\footnotetext{\textsuperscript{e}\,“Yes” indicates that complex post-processing (e.g., high-order differentiation, bit-order reversal, and/or XOR operations) is required. All methods}
\footnotetext{~~marked “Yes” employ a commonly adopted least-significant-bit extraction approach.}
\footnotetext{\textsuperscript{f}\,“No” indicates that no complex post-processing is required prior to bit extraction.}
\footnotetext{\textsuperscript{g}\,Values in parentheses represent random bit generation rates or channel counts claimed as potential or achievable in that study.}

\end{sidewaystable}

\clearpage

\section*{Methods}

\subsection*{Laser configuration and chaos generation}

The comb-laser architecture illustrated in Figs.~\ref{fig1} and~\ref{fig3} comprises a commercially available single-mode distributed feedback (DFB) semiconductor laser, a lithium-niobate optical phase modulator, and a signal generator (see Fig.~1a). The DFB laser was biased at $5.5I_{\rm th}$, where $I_{\rm th}$ denotes the laser threshold current, and thermally stabilized at 19.83$^\circ$C, yielding a laser output power of about 11.5~mW. The signal generator supplied a 6-GHz modulation tone at a power of 14~dBm to drive the phase modulator. Under these operating conditions, the system generated an optical frequency comb with a central frequency of 193.405~THz (1550.08~nm in wavelength), a comb-line spacing of 6~GHz, and a total output power of approximately 6.8~mW.

Four commercially available DFB semiconductor lasers without built-in isolators (Laser Diode~1--4) were employed in the parallel configuration shown in Fig.~\ref{fig3}a, with Laser Diode~1 corresponding to the laser diode used for optical feedback in Fig.~\ref{fig1}. Each laser was biased at approximately $6I_{\rm th}$, where $I_{\rm th}$ denotes the corresponding threshold current. The temperatures of Laser Diode~1--4 were set to 25$^\circ$C, 23.27$^\circ$C, 23.59$^\circ$C, and 20.42$^\circ$C, respectively, resulting in a frequency detuning of $\Delta f=-70$~GHz (0.56~nm in wavelength) relative to the central frequency of the comb laser. The frequency detuning $\Delta f$ is defined as
\begin{equation}
\Delta f \equiv f_{{\rm LD},m} - f_{\rm comb},
\end{equation}
where $f_{\rm comb}$ is the central frequency of the comb laser and $f_{{\rm LD},m}$ is the free-running oscillation frequency of Laser Diode~$m$. Under these conditions, all $f_{{\rm LD},m}$ values ($m=1$--$4$) were approximately 193.335~THz (1550.64~nm in wavelength). The output powers of Laser Diode~1--4 were 10.9, 12.7, 13.3, and 12.8~mW, respectively, with relaxation resonance frequencies of approximately 11~GHz. Each laser output was coupled to an optical feedback loop with round-trip delay times of approximately 53.2, 53.1, 53.8, and 53.6~ns for Laser Diode~1--4, respectively. The feedback strength was quantified as $\xi_{\rm f}=\sqrt{P_{\rm fb}/P_{\rm out}}$, where $P_{\rm fb}$ and $P_{\rm out}$ denote the feedback and free-running output powers, respectively. The values of $\xi_{\rm f}$ for Laser Diode~1--4 were 0.55, 0.55, 0.35, and 0.34, respectively. Each WCBES unit in Fig.~\ref{fig3}a includes the comb-laser output, the emission from the corresponding chaotic laser, and a high-speed photodetector. The detailed specifications are provided in Supplementary Note~I.

\subsection*{Signal acquisition and spectral–temporal analysis}

Chaotic signal intensities were detected using a high-speed photodetector with a 3-dB bandwidth of 100~GHz and digitized using a real-time oscilloscope with a 110-GHz analog bandwidth and a 256-GS/s sampling rate. Electrical spectra were computed using the MATLAB \texttt{pspectrum} function, with Kaiser windowing ($\beta = 6$) applied to time-series segments containing 10 million data points. The standard bandwidth~\cite{lin2012effective} is defined as the frequency span from DC to the point at which 80\% of the total spectral power is accumulated. Spectral flatness~\cite{johnston2002transform} was calculated as the ratio of the geometric mean to the arithmetic mean of the spectral power within this frequency range. Spectral flatness ranges from 0 to 1, with values approaching 1 indicating a flatter and more uniform spectral distribution. In addition, the effective bandwidth~\cite{lin2012effective} is defined as the sum of the largest discrete spectral segments of the electrical spectrum until 80\% of the total spectral power is accumulated.

Temporal characteristics were evaluated via autocorrelation analysis:
\begin{equation}
C_v = \frac{\langle(I_t - \langle I_t \rangle)(I_{t+v} - \langle I_t \rangle)\rangle}{\langle(I_t - \langle I_t \rangle)^2\rangle},
\end{equation}
where $I_t$ represents the time-series data of the entropy source, $\langle I_t \rangle$ is the temporal mean, $v$ is the time-lag index, and $\langle \cdot \rangle$ denotes the ensemble average. The autocorrelation coefficient $C_v$ ranges from $-1$ to $1$, providing a normalized measure of temporal correlation at different delays. For the autocorrelation analysis shown in Fig.~\ref{fig1}e, 10~million data samples were analyzed. Similarly, for the bitstream autocorrelation analysis in Fig.~\ref{fig2}c, 10 million bits were evaluated.

Cross-correlations among the four parallel channels were calculated as
\begin{equation}
C_{ij} = \frac{\langle(I_t^{(i)} - \langle I_t^{(i)} \rangle)(I_{t}^{(j)} - \langle I_t^{(j)} \rangle)\rangle}{\sqrt{\langle(I_t^{(i)} - \langle I_t^{(i)} \rangle)^2\rangle\langle(I_t^{(j)} - \langle I_t^{(j)} \rangle)^2\rangle}},
\end{equation}
where $I_t^{(i)}$ and $I_t^{(j)}$ represent the time-series data from the $i$-th and $j$-th channels ($i,j = 1$–$4$). The cross-correlation coefficient $C_{ij}$ ranges from $-1$ to $1$, and quantifies the similarity and statistical dependence between channel pairs. For the bitstream cross-correlation analysis in Fig.~\ref{fig3}d, 10 million binary bits per channel were evaluated.

\subsection*{Entropy rate estimation and random bit generation}

For parallel RBG, four channels (Keysight UXR1104B) were used, each operating as a 10-bit ADC with a 110 GHz analog bandwidth and a 256 GS/s sampling rate. Hardware limitations confined the demonstration to four channels.

Entropy rate estimation followed the minimum-entropy procedure recommended by NIST SP~800-90B~\cite{NIST.SP.800-90B-2018} for physical entropy sources. Entropy per sample was first evaluated, and the entropy rate was obtained by multiplying this value by the sampling rate. Because the chaos-based entropy source does not satisfy the independent and identically distributed (i.i.d.) assumption, the non-i.i.d. entropy estimators specified in NIST SP~800-90B were considered, including the most common value (MCV), collision, Markov, compression, t-tuple, longest repeated substring (LRS), multi most common in window (MultiMCW) prediction, lag prediction, multiple Markov models with counting (MultiMMC) prediction, and LZ78Y prediction. However, because the collected data samples in this study consist of 10-bit binary data, the collision, Markov, and compression estimators, which are applicable only to 1-bit binary data, were not applied. The remaining seven estimators were used for the entropy assessment~\cite{tseng2024entropy}. 

For entropy measurement, eleven independent trials were performed, each consisting of 1 million data samples acquired at different time points under identical experimental conditions. For each trial, all seven entropy estimators were applied to the same dataset, and the minimum among the seven resulting estimates was taken as the entropy value for that trial. The median of these eleven minimum-entropy values was reported as the source entropy and was further validated using the restart test. The entropy rate, denoted $H_{\rm NIST}$, was obtained by multiplying the validated source entropy by the sampling rate of 256~GS/s.

For validation, results were compared with the theoretical upper limit defined by the Shannon–Hartley theorem~\cite{hart2017recommendations,tseng2024entropy}:
\begin{equation}
H_{\rm SH} = \min(2BW,f_s)\times \left(n - \sum_{x} p(x) \log_2\!\left(\frac{p(x)}{u(x)}\right)\right),
\end{equation}
where $BW$ is the ADC analog bandwidth (110~GHz), $f_s$ is the sampling rate (256~GS/s), $n$ is the number of bits per sample, $x$ denotes the digitized sample value, $p(x)$ is the probability distribution of the entropy source, and $u(x)$ is a uniform distribution over the support of $p(x)$.

Bitstream quality was evaluated using the NIST SP~800-22 statistical test suite~\cite{NIST.SP.800-22rev1a-2010}, which comprises 15 independent tests, each detecting distinct types of nonrandom patterns (Fig.~\ref{fig2}f). In most previous studies~\cite{uchida2008fast,kanter2010optical,kim2021massively,shen2023harnessing}, NIST SP~800-22 was applied to 1000 sequences of 1-Mbit data, with a significance level of 0.01 for a single trial. Under these conditions, a test in the NIST SP~800-22 suite passes if the $P$-value (uniformity of $p$-values) exceeds 0.0001, and the pass proportion lies within $0.99 \pm 0.0094$. A sequence passes all 15 tests to satisfy the NIST SP~800-22 criteria for a single trial.
To meet these criteria and enhance reliability under a more conservative standard, we evaluated the bitstream generated by the entropy source across eleven independent NIST SP~800-22 trials. Trials were performed on different 1-Gbit sequences obtained at different time points under identical experimental conditions. This procedure verified that the randomness performance was consistent and repeatable across independent sequences. In this study, a bitstream had certified randomness if the median number of passed NIST tests was 15 for the eleven trials.

\bmhead{Acknowledgements}
We express our gratitude to Keysight Technologies in Taiwan for their support in the measurements. This work was supported in part by JSPS KAKENHI (JP22H05195, JP25H01129, and JP25KF0127), JST CREST (JPMJCR24R2), and the National Science and Technology Council, Taiwan (NSTC 113-2112-M-006-007, 114-2112-M-006-004, and 113-2221-E-006-110-MY3).

\bmhead{Author contributions}
C.-H. Tseng conceived the idea, designed and performed the experiments, analyzed the data, and wrote the manuscript, with conceptual aspects of the work discussed with S.-K. Hwang. A. Uchida and S.-K. Hwang contributed to discussions on the manuscript structure, provided valuable suggestions during the manuscript revision, and supervised the project.

\bmhead{Competing interests}
The authors declare no competing interests.

\bmhead{Data availability}
The data that support the findings of this study are available from the corresponding authors upon reasonable request.

\bmhead{Corresponding authors}
Correspondence and requests for materials should be addressed to C.-H. Tseng or S.-K. Hwang.


\section*{Supplementary Information}

\setcounter{figure}{0}
\setcounter{table}{0}
\setcounter{equation}{0}

\renewcommand{\figurename}{Supplementary Fig.}
\renewcommand{\tablename}{Supplementary Table}
\renewcommand{\theequation}{S\arabic{equation}}

\setcounter{topnumber}{2}
\setcounter{bottomnumber}{1}
\setcounter{totalnumber}{3}
\renewcommand{\topfraction}{0.9}
\renewcommand{\bottomfraction}{0.8}
\renewcommand{\textfraction}{0.07}
\renewcommand{\floatpagefraction}{0.9}

\setlength{\abovecaptionskip}{4pt}   
\setlength{\belowcaptionskip}{2pt}   
\setlength{\intextsep}{8pt}          
\setlength{\textfloatsep}{10pt}      
\setlength{\floatsep}{8pt}           

\subsection*{Supplementary Note I: Detailed experimental configuration}

\begin{figure}[H]
\includegraphics[width=1.0\linewidth]{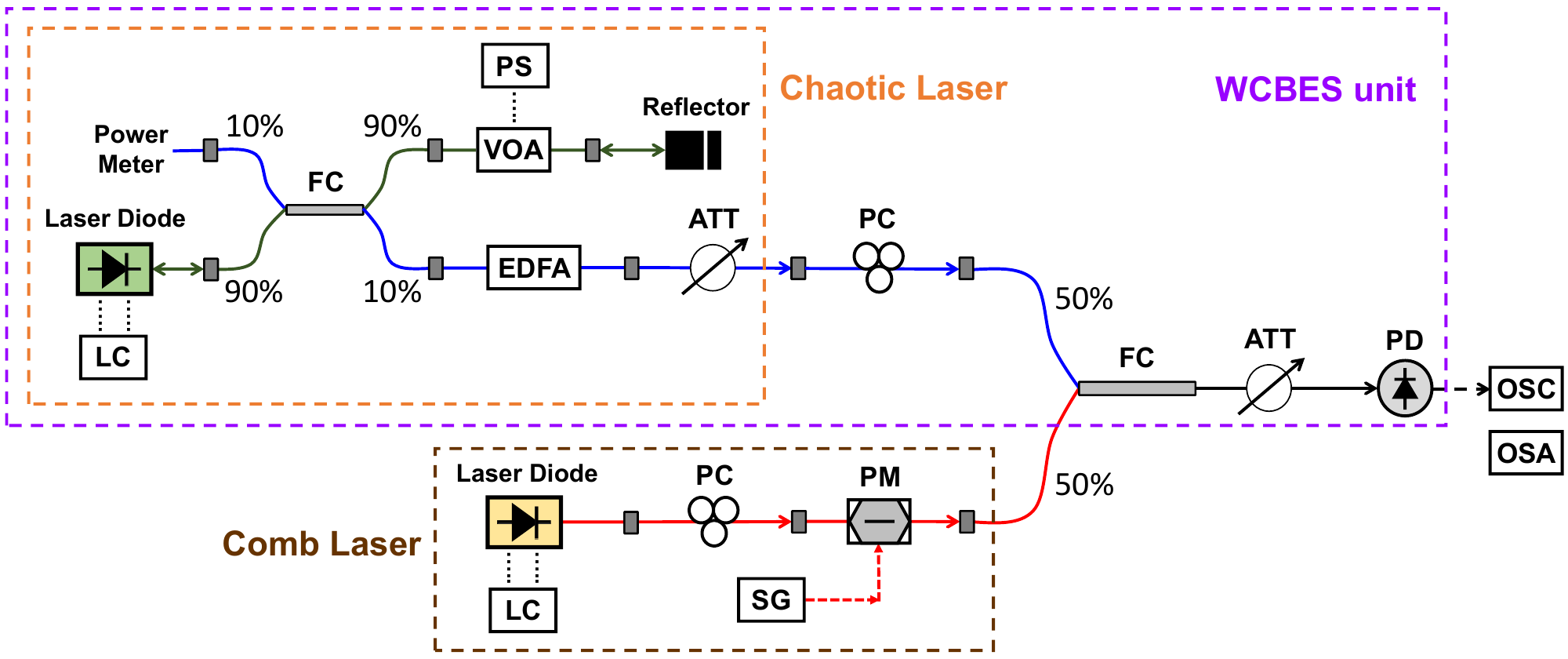}
\caption{\textbf{Detailed schematic of the experimental setup.}\\
ATT, variable optical attenuator; EDFA, erbium-doped fiber amplifier; FC, fiber coupler;
LC, laser controller; OSA, optical spectrum analyzer; OSC, real-time digital oscilloscope;
PC, polarization controller; PD, photodetector; PM, power meter; PS, power supply;
SG, signal generator; VOA, voltage-controlled variable optical attenuator.}
\end{figure}

In the experimental setup (Supplementary Fig.~1), the laser diode for chaos generation (G\&H, AA0702) and the laser diode for comb generation (Furukawa, FRL15DCWD\-A81) are packaged in butterfly modules with integrated thermoelectric coolers and monitoring photodiodes, each equipped with a 1-m fiber pigtail. The temperatures and bias currents of both lasers are stabilized using high-resolution laser controllers (Newport, 8610.8C). The optical feedback loop comprises a 2\texttimes2, 90\%:10\% fiber coupler (Thorlabs, PNH1550R2A2), an in-line voltage-controlled variable optical attenuator (Thorlabs, V1550PA), and a fiber reflector (Thorlabs, P5-1550PMR-P01-1). The feedback strength is controlled by the voltage-controlled variable optical attenuator, and the feedback power is monitored in real time using a power meter (Thorlabs, PM20A). All components in the feedback loop are polarization-maintaining. An erbium-doped fiber amplifier (Amonics, AEDFA-BO-23-B-FA) and a variable optical attenuator in the blue path are used to adjust the optical output power from the 2\texttimes2 fiber coupler so that it matches the output power of the comb laser. A polarization controller in the blue path ensures polarization matching between the blue and red optical paths.

The optical frequency comb is generated by phase modulating the output of a separate laser diode (Furukawa, FRL15DCWD-A81) using a lithium-niobate optical phase modulator. A polarization controller in the red path aligns the polarization state of the laser diode output field with the $Z$-cut axis of the lithium-niobate phase modulator (EOspace, PM-DV5-40). The modulator exhibits a half-wave voltage of $V_{\pi} = 3.9$~V at 1~GHz and a 3-dB bandwidth of approximately 30~GHz. A low-noise electrical signal generator (Agilent, E8257D) supplies a 6~GHz sinusoidal waveform with a power level of 14~dBm to drive the phase modulator. The optical powers in the blue and red paths are then combined in equal proportion using a 2\texttimes1, 50\%:50\% fiber coupler.

\begin{table}[t]
\renewcommand{\arraystretch}{1.3}
\noindent\caption{Summary of Equipment and Component Specifications}

\vspace{0em}

\centering
\begin{tabular}{|m{3.3cm}|m{3.8cm}|m{4.2cm}|}
\hline
\textbf{Component} & \textbf{Manufacturer \newline and Model} & \textbf{Key Specifications} \\
\hline
Laser Diode (1--4) for chaos generation & G\&H, AA0702 with fiber pigtail (no isolator) & Optical frequency: 193.335~THz \newline bias current: 73~mA ($6I_{\rm th}$) \\
\hline
Laser Diode for \newline comb generation & Furukawa, FRL15DCWD-A81 with fiber pigtail & Optical frequency: 193.405~THz \newline bias current: 60~mA ($5.5I_{\rm th}$) \\
\hline
Laser controller & Newport, 8610.8C & Current~stability: $<$~10~ppm~SF Temperature stability: \newline $\pm$~0.0005$^\circ$C\\
\hline
Optical phase \newline modulator & EOspace, PM-DV5-40  & $V_{\pi} = 3.9$~V at 1~GHz \newline 3-dB Bandwidth: 30 GHz\\
\hline
2\texttimes2, 90\%:10\% split \newline fiber coupler & Thorlabs, PNH1550R2A2 & Polarization-maintaining\\
\hline
Voltage-controlled \newline variable optical \newline attenuator & Thorlabs, V1550PA  & Polarization-maintaining\\
\hline
Reflector & Thorlabs, \newline P5-1550PMR-P01-1 & Polarization-maintaining\\
\hline
Erbium-doped \newline fiber amplifier & Amonics, \newline AEDFA-BO-23-B-FA & Maximum gain: 23 dB\\
\hline
Photodetector & Fraunhofer, HHI & 3-dB bandwidth: 100\,GHz \\
\hline
Optical spectrum \newline analyzer & Advantest, Q8384 & Wavelength resolution: 0.01\,nm \newline at 1550~nm\\
\hline
Real-time digital\newline oscilloscope & Keysight, UXR1104B & Analog bandwidth: 110\,GHz \newline Sampling rate: 256\,GS/s \newline ADC resolution: 10 bits \\
\hline
\end{tabular}
\end{table}

The combined optical signal is analyzed using an optical spectrum analyzer (Advantest Q8384) to characterize its spectral features. It is also directed to a high-speed photodetector (Fraunhofer HHI) with a 3-dB bandwidth of 100~GHz and then measured using a real-time digital oscilloscope (Keysight UXR1104B) with an analog bandwidth of 110~GHz and a sampling rate of 256~GS/s to evaluate the temporal and spectral properties of the wideband chaos-based entropy source (WCBES). The input optical power to the photodetector is set to approximately 4~mW to ensure a sufficiently high signal-to-noise ratio. The WCBES unit shown in Fig.~3 of the main text corresponds to the region enclosed by the purple dashed box in Supplementary Fig.~1. In this study, four parallel WCBES units are demonstrated, each incorporating one laser diode (Laser Diode~1--4). A complete list of equipment and their model numbers is provided in Supplementary Table~1.

\clearpage
\subsection*{Supplementary Note II: Spectral optimization of chaos induced by delayed optical self-feedback}

\begin{figure}[H]
\centering
\includegraphics[width=1.0\linewidth]{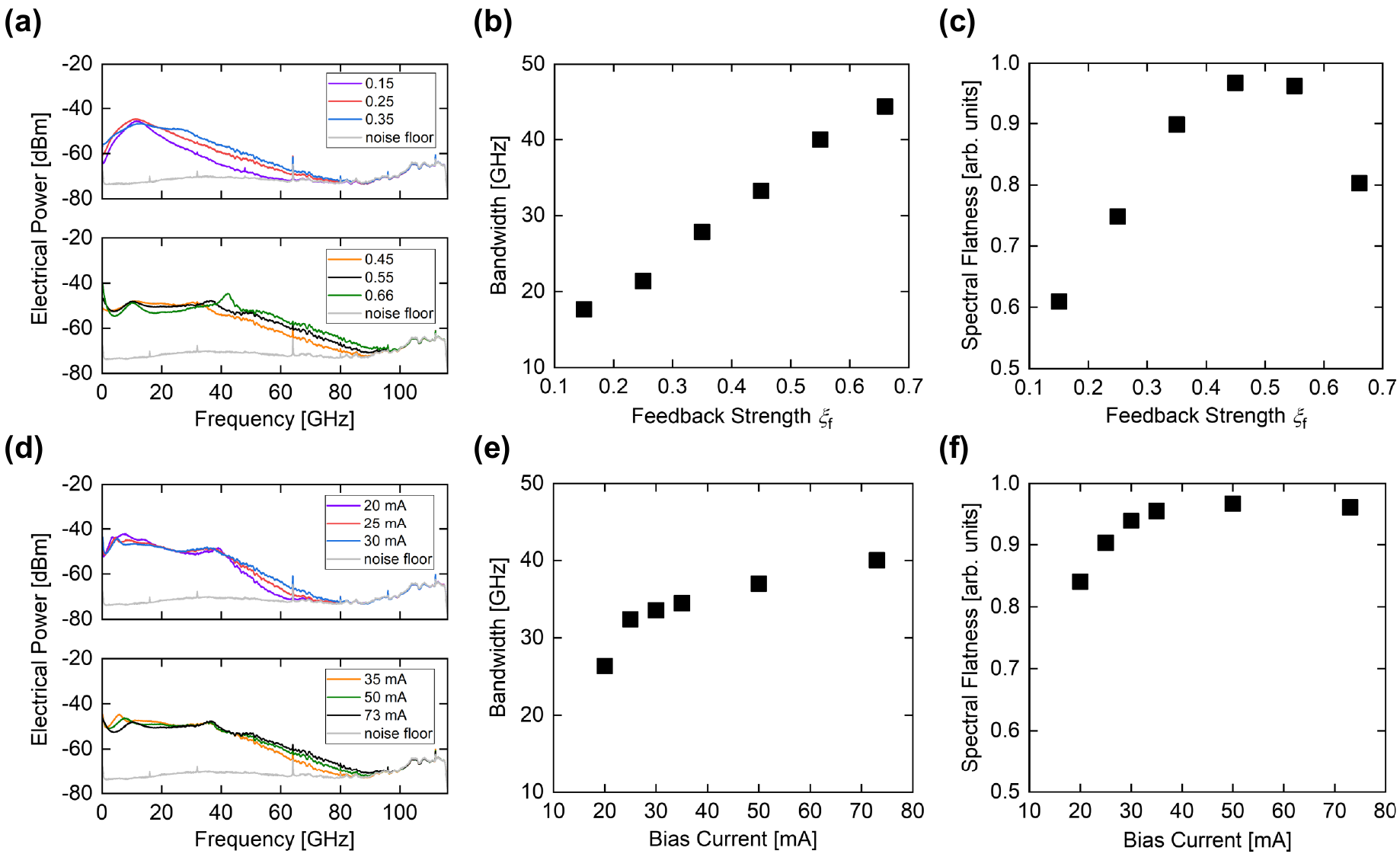}
\caption{\textbf{Spectral characteristics of chaos induced by delayed optical self-feedback.}
\textbf{a,} Chaos spectra for different optical feedback strengths $\xi_{\rm f}$.
\textbf{b,} Standard bandwidth and \textbf{c,} spectral flatness as a function of $\xi_{\rm f}$.
\textbf{d,} Chaos spectra for different bias currents.
\textbf{e,} Standard bandwidth and \textbf{f,} spectral flatness as a function of bias current.}
\end{figure}

In this section, we examine how the optical feedback strength and the bias current of a semiconductor laser influence the spectral characteristics of chaos generated through delayed optical self-feedback. By optimizing these parameters, the chaotic emission from the semiconductor laser attains substantially enhanced spectral properties compared with previously reported demonstrations of optical-feedback-induced chaos in semiconductor lasers, which typically exhibit bandwidths of only several tens of gigahertz and uneven spectral profiles, thereby constraining the achievable entropy throughput~\cite{hart2017recommendations}. Representative results obtained from Laser Diode~1 under delayed optical self-feedback are presented here, while the performances of Laser Diode~1--4 are found to be highly consistent.

We first investigate the effect of varying the optical feedback strength $\xi_{\rm f}$ while keeping the bias current at 73~mA. The feedback strength $\xi_{\rm f}$ is defined as the square root of the power ratio between the optical feedback and the free-running laser output, as described in the Methods section of the main text. Throughout this document, $\xi_{\rm f}$ is set to 0.15, 0.25, 0.35, 0.45, 0.55, and 0.65. As shown in Supplementary Fig.~2a, at lower feedback strengths (e.g., $\xi_{\rm f} = 0.15$ and 0.25), the chaos spectrum is dominated by a prominent peak near the relaxation oscillation frequency of the semiconductor laser. This behavior is consistent with earlier studies~\cite{uchida2008fast} and reflects a commonly observed regime of optical-feedback-induced chaos in semiconductor lasers. As the feedback strength increases, both the bandwidth and the uniformity of the spectral power distribution improve steadily. However, when $\xi_{\rm f}$ exceeds 0.55, a degradation in the spectral uniformity emerges, indicating that the optimal feedback regime is reached near $\xi_{\rm f} = 0.55$, which we adopt in our experiments. The corresponding standard bandwidth and spectral flatness, both defined in the Methods section of the main text, are plotted as functions of $\xi_{\rm f}$ in Supplementary Figs.~2b and 2c, respectively. At $\xi_{\rm f} = 0.55$, the standard bandwidth expands to 40~GHz, and the spectral flatness attains 0.96, marking the optimal operating point for producing broadband and spectrally uniform chaos from the chaotic laser (see Fig.~1c, blue curve, in the main text).

We next fix the feedback strength at $\xi_{\rm f} = 0.55$ and examine the influence of the bias current on the resulting chaos spectra. The bias current is swept from 20 to 73~mA (20, 25, 30, 35, 50, and 73~mA). As shown in Supplementary Fig.~2d, increasing the bias current shifts the relaxation oscillation frequency upward and produces progressively flatter spectra, thereby broadening the chaos bandwidth and improving the spectral flatness. The corresponding standard bandwidth and spectral flatness, plotted as functions of the bias current, are shown in Supplementary Figs.~2e and 2f, respectively. To avoid operating close to the maximum allowable bias current of the laser diode (80~mA), which could lead to device degradation or damage, we limit the maximum bias current to 73~mA and adopt this value as the optimized operating point.

Through careful optimization of both the bias current and the feedback strength, we achieve fast chaotic emission from the laser diode, yielding a standard bandwidth of 40~GHz and a spectral flatness of 0.96. By combining this chaotic emission with an optical frequency comb and performing optical heterodyning using a high-speed photodetector, the standard bandwidth of the resulting chaos-based entropy source is extended to 104~GHz in our experiment (limited by the measurement bandwidth of the equipment) and is theoretically expected to exceed 150~GHz. The underlying mechanism is elucidated through numerical simulations in the following section.

\clearpage
\subsection*{Supplementary Note III: Spectral broadening mechanism}
\subsubsection*{1. A semiconductor laser with delayed optical self-feedback}

Before explaining the spectral broadening mechanism, we first describe the dynamical behavior of a semiconductor laser subject to optical feedback using the Lang–Kobayashi rate equations~\cite{uchida2012optical}.

\begin{equation}
\begin{aligned}
\frac{dE_{\rm LD}(t)}{dt} &= \frac{1+i\alpha}{2}\Bigg[ \frac{G_N \big(N_{\rm LD}(t) - N_{0}\big)}{1+\varepsilon |E_{\rm LD}(t)|^2} - \frac{1}{\tau_{p}}\Bigg]E_{\rm LD}(t) + \beta(t) \\
&\quad + \frac{1}{\tau_{\mathrm{in}}}\frac{r_2^2-1}{r_2^2} \sum_{k=1}^{Q} (-r_2r_3)^k E_{\rm LD}(t-k\tau) \exp(-i\omega_{\rm LD} k\tau), \\
\frac{dN_{\rm LD}(t)}{dt} &= J_{\rm LD} - \frac{N_{\rm LD}(t)}{\tau_{s}}
- \frac{G_N \big(N_{\rm LD}(t) - N_{0}\big)}{1+\varepsilon |E_{\rm LD}(t)|^2}|E_{\rm LD}(t)|^2,
\end{aligned}
\label{eq:lk}
\end{equation}

\noindent
$E_{\rm LD}(t)$ and $N_{\rm LD}(t)$ denote the complex intracavity field amplitude and the carrier density of the laser diode used for optical feedback, respectively. The parameters $G_N$ and $N_0$ represent the differential gain coefficient and the carrier density at transparency; $\varepsilon$ is the gain saturation coefficient; $\omega_{\rm LD}$ is the laser angular oscillation frequency; $\alpha$ is the linewidth enhancement factor; and $\tau_p$ and $\tau_s$ correspond to the photon and carrier lifetimes, respectively. The parameter $J_{\rm LD}$ denotes the bias current density; $\tau_{\mathrm{in}}$ is the round-trip time of the internal cavity; $r_2$ is the facet reflectivity of the laser; $r_3$ is the reflectivity of the external mirror; and $\tau$ is the round-trip time of the feedback light in the external cavity. Under strong optical feedback, multiple delayed feedback terms must be included. We set the number of multiple reflections to $Q = 7$ in the simulations. The laser parameters used in the simulations are summarized in Supplementary Table~2.

The rate equations are solved numerically using the fourth-order Runge–Kutta method with a simulation time step of 0.5~ps and a total integration time of 10~$\mu$s. A white Gaussian noise term, $\beta(t)$, is included to simulate spontaneous emission noise and is adjusted such that the laser linewidth is approximately 600~kHz, consistent with that of the laser diode used in the experiment. The simulated chaotic optical spectrum of the laser diode, obtained by taking the Fourier transform of $E_{\rm LD}(t)e^{i\omega_{\rm LD} t}$, is shown in Supplementary Fig.~3a. Throughout the simulations, the frequency axes of the optical spectra are referenced to $\omega_{\rm LD}/2\pi$. The simulation result closely reproduces the chaotic component presented in Fig.~1d of the main text.

The electrical spectrum of the chaotic intensity, $I_{\rm LD}(t) = |E_{\rm LD}(t)|^2$, is shown in Supplementary Fig.~3b, exhibiting a standard bandwidth of 39~GHz and a spectral flatness of 0.97, in good agreement with the experimental result shown in Fig.~1c (upper blue trace) of the main text. Throughout the simulations, the responsivity of the photodetector is assumed to be 1. The autocorrelation trace of $I_{\rm LD}(t)$ is presented in Supplementary Fig.~3c, where the power of the strongest side peak (time-delay signature) is approximately 0.44, which is on the same order as the experimental observation in Fig.~1e (upper blue trace) of the main text. This correlation peak should be suppressed for random number generation applications. To achieve this, we combine the chaotic emission from the laser diode with an optical frequency comb, where optical heterodyning between the two generates a wideband chaos-based entropy source (WCBES) with a strongly suppressed time-delay signature. The following subsection describes the optical frequency comb generation process in detail.

\begin{table}[h]
\centering
\caption{Laser parameters used for numerical simulation of the Lang--Kobayashi equations.}
\label{tab:params}
\renewcommand{\arraystretch}{1.2}
\small  
\begin{tabular}{lll}
\toprule
\textbf{Symbol} & \textbf{Parameter} & \textbf{Value} \\
\midrule
$G_N$       & Gain coefficient & $5.88 \times 10^{-12}~\mathrm{m^3\,s^{-1}}$ \\
$N_0$       & Carrier density at transparency & $7.0 \times 10^{23}~\mathrm{m^{-3}}$ \\
$\epsilon$       & Gain saturation coefficient & $2.0 \times 10^{-23}$ \\
$\tau_p$    & Photon lifetime & $1.927 \times 10^{-12}~\mathrm{s}$ \\
$\tau_s$    & Carrier lifetime & $2.04 \times 10^{-9}~\mathrm{s}$ \\
$\tau_{\mathrm{in}}$ & Round-trip time in internal cavity & $8.0 \times 10^{-12}~\mathrm{s}$ \\
$r_2$       & Reflectivity of laser facet & $0.556$ \\
$r_3$       & Reflectivity of external mirror & $0.9$ \\
$j_{\rm LD} = J_{\rm LD}/J_{\mathrm{th}}$ & Normalized bias current density & $6.0$ \\
$\alpha$    & Linewidth enhancement factor & $3.6$ \\
$\tau$ & Delay time of optical feedback & $53.22~\mathrm{ns}$ \\
$\lambda$   & Optical wavelength & $1550.64~\mathrm{nm}$ \\
$c$         & Speed of light & $2.998 \times 10^{8}~\mathrm{m\,s^{-1}}$ \\
$N_{\mathrm{th}} = N_0 + \dfrac{1}{G_N \tau_p}$ & Carrier density at threshold & $7.883 \times 10^{23}~\mathrm{m^{-3}}$ \\
$J_{\mathrm{th}} = N_{\mathrm{th}}/\tau_s$ & bias current density at threshold & $3.864 \times 10^{32}~\mathrm{m^{-3}\,s^{-1}}$ \\
$\omega_{\rm LD} = \dfrac{2\pi c}{\lambda}$ & Laser angular oscillation frequency & $1.215 \times 10^{15}~\mathrm{s^{-1}}$ \\
\bottomrule
\end{tabular}
\end{table}

\begin{figure}[H]
    \centering
    \includegraphics[width=1.0\linewidth]{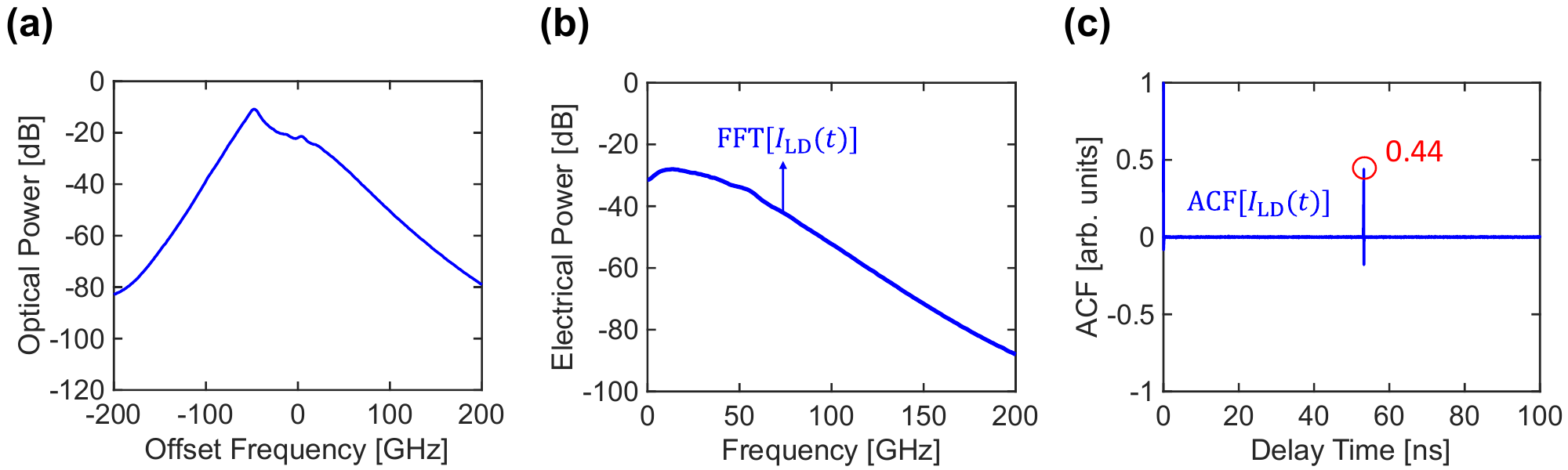}
    \caption{\textbf{Simulation results of chaos induced by delayed optical self-feedback.}
    \textbf{a,} Simulated optical spectrum of the intracavity field $E_{\rm LD}(t)e^{i\omega_{\rm LD} t}$. 
    \textbf{b,} Simulated electrical spectrum of the chaos intensity $I_{\rm LD}(t) = |E_{\rm LD}(t)|^2$. 
    \textbf{c,} Autocorrelation trace of $I_{\rm LD}(t)$, showing a strong time-delay signature.}
\end{figure}

\clearpage
\subsubsection*{2. Phase modulation for optical frequency comb generation}

The output field of the phase-modulated laser diode is expressed as
\begin{equation}
E_{\mathrm{PM}}(t) = E_0 \exp\Big[i\big(\omega_{\rm c} t + \gamma \sin(\omega_{\rm m} t)\big)\Big],
\end{equation}
where $E_0$ is the constant field amplitude of the laser diode used for comb generation, $\omega_{\rm c}$ is its angular oscillation frequency, $\omega_{\rm m} = 2\pi f_{\rm m}$ is the angular modulation frequency, and $\gamma$ is the modulation index. For simplicity, the spontaneous emission noise of the laser diode used for comb generation is not included in the simulations. The frequency detuning between the laser diode used for comb generation and the laser diode used for chaos generation is given by
\[
\frac{\omega_{\rm i}}{2\pi} = \frac{\omega_{\rm c} - \omega_{\rm LD}}{2\pi} = 70~\mathrm{GHz}.
\]

\noindent
For a LiNbO$_3$ phase modulator with a $50~\Omega$ impedance at the electrical input port, the modulation index $\gamma$ is defined as
\[
\gamma = \pi \frac{V_{\mathrm{pk}}}{V_{\pi}}, \qquad
V_{\mathrm{pk}} = \sqrt{2}\,V_{\mathrm{rms}}, \qquad
V_{\mathrm{rms}} = \sqrt{P_{\mathrm{RF}} R}.
\]
Here, $V_{\pi}$ is the half-wave voltage of the modulator, $V_{\mathrm{rms}}$ is the root-mean-square voltage corresponding to the electrical modulation power $P_{\mathrm{RF}}$, and $V_{\mathrm{pk}}$ is the peak voltage of the electrical modulation signal. In our setup, the following parameter values are used:
\[
f_{\rm m} = 6~\mathrm{GHz}, \quad
P_{\mathrm{RF}} = 14~\mathrm{dBm}~(25~\mathrm{mW}), \quad
R = 50~\Omega, \quad
V_{\pi} = 3.9~\mathrm{V}.
\]

\noindent
These values yield
\[
V_{\mathrm{rms}} \approx 1.118~\mathrm{V}, \quad
V_{\mathrm{pk}} \approx 1.581~\mathrm{V}, \quad
\gamma \approx 1.27.
\]

\medskip
\noindent
Applying the Jacobi--Anger expansion, where $J_n(\gamma)$ represents the Bessel function of the first kind of order $n$,
\begin{equation}
e^{i\gamma \sin(\omega_{\rm m} t)} = \sum_{n=-\infty}^{\infty} J_n(\gamma)\, e^{i n \omega_{\rm m} t},
\end{equation}
the phase-modulated optical field becomes
\begin{equation}
E_{\mathrm{PM}}(t) = E_0 \sum_{n=-\infty}^{\infty} J_n(\gamma)\, e^{i(\omega_{\rm c} + n\omega_{\rm m})t}.
\end{equation}

\noindent
This expression indicates that the phase-modulated output forms an optical frequency comb with spectral lines at $\omega_{\rm c} \pm n\omega_{\rm m}$, each weighted by the corresponding coefficient $J_n(\gamma)$. In the following subsection, we demonstrate optical heterodyning between the chaotic emission $E_{\rm LD}(t)e^{i\omega_{\rm LD} t}$ and the optical frequency comb $E_{\mathrm{PM}}(t)$.

\vspace{1ex}

\subsection*{3. Heterodyning between the chaotic emission and the optical frequency comb}

To suppress the time-delay signature and broaden the spectrum, the chaotic emission,
$E_{\rm LD}(t)e^{i\omega_{\rm LD} t}$, is combined with the optical frequency comb,
$E_{\mathrm{PM}}(t)$, and directed to a high-speed photodetector. For consistency with the experimental conditions, the powers of $E_{\rm LD}(t)e^{i\omega_{\rm LD} t}$ and $E_{\mathrm{PM}}(t)$ are normalized to be equal. The optical spectrum of the combined field,
$E_{\rm LD}(t)e^{i\omega_{\rm LD} t}+E_{\mathrm{PM}}(t)$, is shown in Supplementary Fig.~4a and agrees well with the experimental result in Fig.~1d of the main text. The detected photocurrent is proportional to the optical intensity of the combined field:
\begin{equation}
I_{\rm w}(t) = \left| E_{\rm LD}(t)e^{i\omega_{\rm LD} t} + E_{\mathrm{PM}}(t) \right|^2.
\end{equation}

\noindent
Expanding this expression yields
\begin{equation}
I_{\rm w}(t) = |E_{\rm LD}(t)|^2 + |E_{\mathrm{PM}}(t)|^2 
+ 2\,\mathrm{Re}\!\left[ E_{\mathrm{PM}}(t) E_{\rm LD}^*(t)e^{-i\omega_{\rm LD} t} \right].
\end{equation}

\noindent
The first term corresponds to the intensity of the chaotic laser,
$I_{\rm LD}(t)=|E_{\rm LD}(t)|^2$.
The second term, $|E_{\mathrm{PM}}(t)|^2 = |E_0|^2$, is constant in time and therefore does not produce comb peaks in the electrical spectrum. This explains the absence of discrete spectral peaks in the electrical spectrum shown in Fig.~1c (lower black trace) of the main text.
The final term represents the heterodyne photomixing between the chaotic field and the optical frequency comb. Substituting Eq.~(S4), the heterodyne term becomes
\begin{equation}
I_{\rm h}(t) = 2\,\mathrm{Re}\!\left[
E_0 E_{\rm LD}^*(t)
\sum_{n=-\infty}^{\infty}
J_n(\gamma)\,
e^{i(\omega_{\rm i} + n\omega_{\rm m})t}
\right].
\end{equation}

\noindent
This expression indicates that the intracavity field amplitude $E_{\rm LD}(t)$ not only undergoes frequency inversion (due to complex conjugation) but is also shifted and replicated around each frequency component $\omega_{\rm i} \pm n\omega_{\rm m}$, with relative weights determined by $J_n(\gamma)$. 
The resulting WCBES, denoted as $I_{\rm w}(t)$, is obtained by combining the chaotic intensity $I_{\rm LD}(t)$ with the heterodyne contribution $I_{\rm h}(t)$.

\begin{figure}[t]
    \centering
    \includegraphics[width=1.0\linewidth]{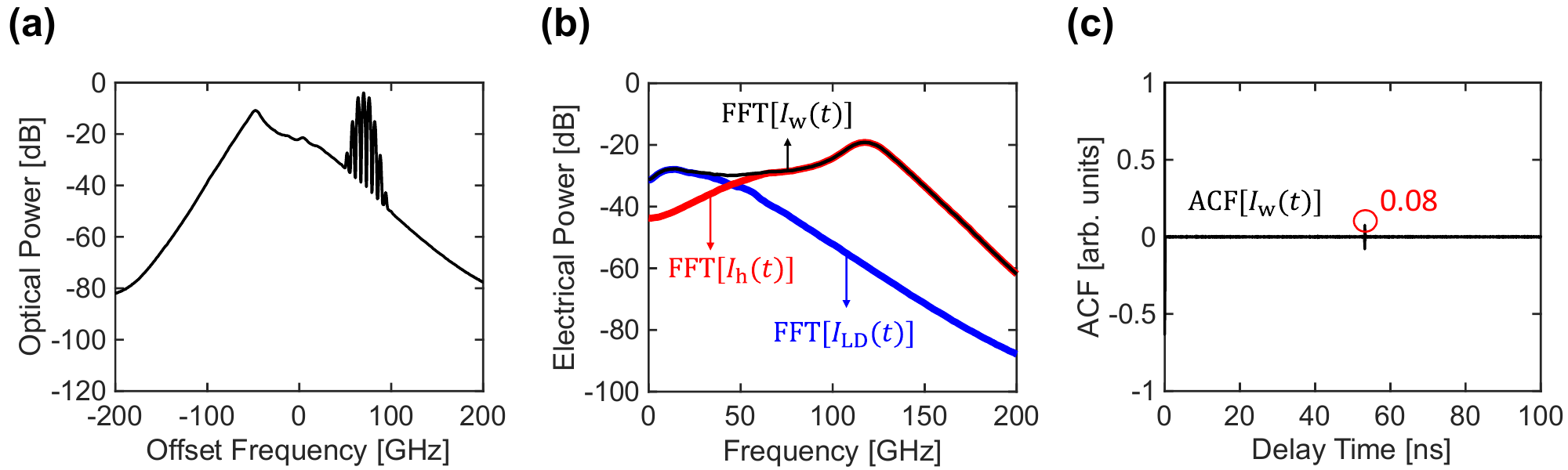}
    \caption{\textbf{Simulation results of wideband chaos-based entropy source (WCBES) generation.} 
    \textbf{a,} Simulated optical spectrum of $E_{\rm LD}(t)e^{i\omega_{\rm LD} t}+E_{\mathrm{PM}}(t)$. 
    \textbf{b,} Simulated electrical spectra of $I_{\rm LD}(t)$ (blue), $I_{\rm h}(t)$ (red), and $I_{\rm w}(t)$ (black). 
    \textbf{c,} Autocorrelation trace of $I_{\rm w}(t)$, showing suppression of the time-delay signature.}
\end{figure}

The Fourier transforms of $I_{\rm LD}(t)$, $I_{\rm h}(t)$, and $I_{\rm w}(t)$ are shown by the blue, red, and black curves, respectively, in Supplementary Fig.~4b. 
The baseband spectrum ($\le 40$~GHz) of $I_{\rm w}(t)$ is dominated by $I_{\rm LD}(t)$, whereas its high-frequency spectrum ($\ge 60$~GHz) arises primarily from $I_{\rm h}(t)$. 
The intermediate frequency region (40--60~GHz) results from the mixing of $I_{\rm LD}(t)$ and $I_{\rm h}(t)$.
Importantly, because the chaotic laser operates under strong optical feedback, $E_{\rm LD}(t)$ already exhibits a broad intrinsic bandwidth.
Through heterodyne photomixing with the optical frequency comb, $E_{\rm LD}^*(t)$ is frequency-shifted and replicated around $\omega_{\rm i} \pm n\omega_{\rm m}$, giving rise to the wideband high-frequency spectral components observed in $I_{\rm h}(t)$ (red curve).
In this process, fine chaotic phase variations of the chaotic field are effectively mapped onto ultrafast intensity fluctuations.

As a result, the spectrum of $I_{\rm w}(t)$ remains highly flat from 0 to 110~GHz, consistent with Fig.~1c (lower black trace) of the main text.
Beyond the bandwidth limitation of the measurement system, numerical simulations further show that the spectrum of $I_{\rm w}(t)$ extends beyond 150~GHz, as illustrated in Supplementary Fig.~4b, confirming the effectiveness of the proposed approach for broadband entropy-source generation.

The autocorrelation trace of $I_{\rm w}(t)$ is shown in Supplementary Fig.~4c.
Owing to wave mixing of $E_{\rm LD}^*(t)$ at frequencies $\omega_{\rm i} \pm n\omega_{\rm m}$ in the $I_{\rm h}(t)$ term, together with mixing between $I_{\rm LD}(t)$ and $I_{\rm h}(t)$, the deterministic structure induced by external feedback is substantially degraded.
As a result, the time-delay signature is strongly suppressed, falling below 0.1 in this case, in agreement with the experimental results shown in Fig.~1e (lower black trace) of the main text.
The simulations, therefore, provide a theoretical explanation for the spectral-broadening mechanism and further demonstrate the effectiveness of the proposed approach for broadband entropy-source generation.

\begin{figure}[t]
    \centering
    \includegraphics[width=1\linewidth]{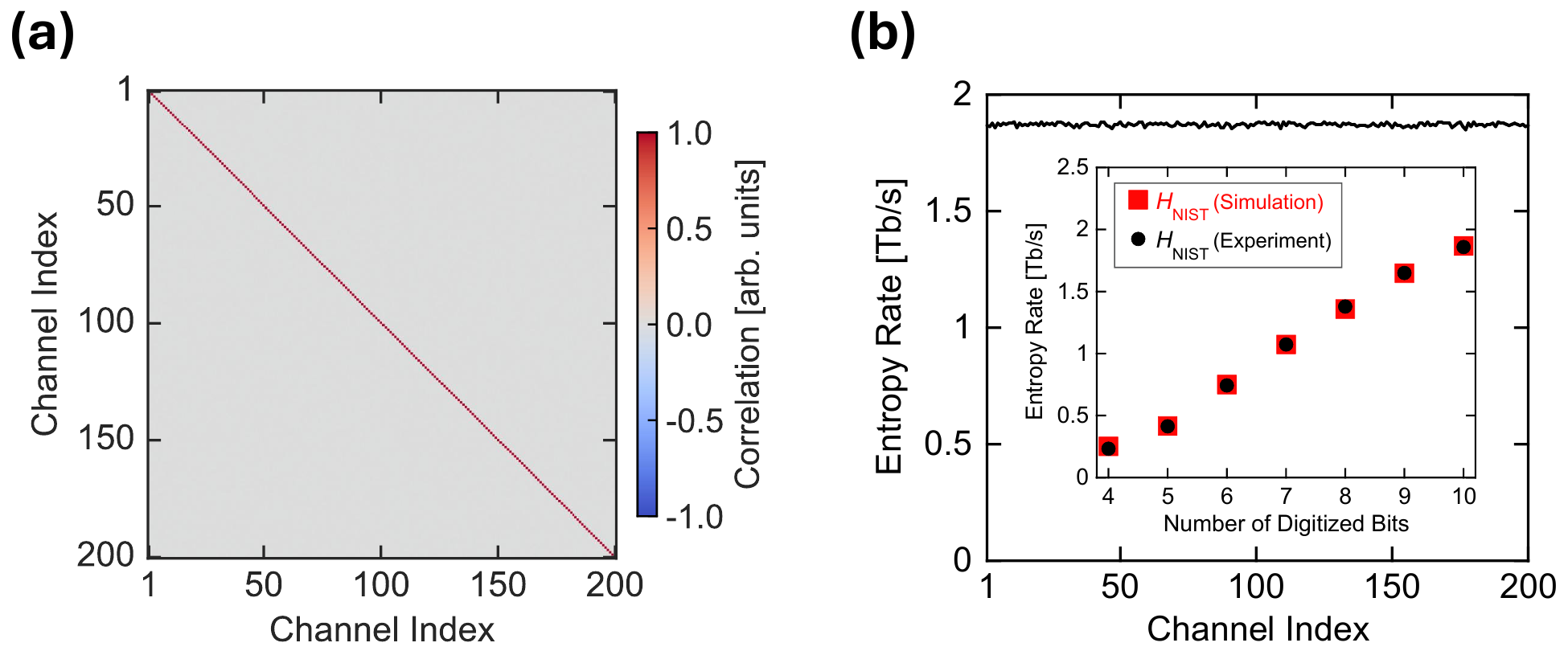}
    \caption{\textbf{Cross-correlation and entropy-rate evaluation for $M$ independent WCBESs.} 
    \textbf{a,} Cross-correlation matrix for $M = 200$ independent WCBES channels.
    \textbf{b,} Estimated entropy rates for 200 channels using the NIST SP~800-90B suite.
    The inset of \textbf{b} shows the average value of $H_{\rm NIST}$ for $I_{\rm w}^{(1)}(t)$ to $I_{\rm w}^{(200)}(t)$ (red symbols) and the experimental $H_{\rm NIST}$ (black symbols) as functions of the number of digitized bits.}
\end{figure}

Finally, we consider the use of $M$ independent laser diodes to further extend the overall entropy-generation capacity, as illustrated in Fig.~3a of the main text.
For simplicity, each chaotic laser diode in the simulations is assumed to operate under identical parameters and conditions while possessing an independent spontaneous-emission noise seed with equal noise power.
Each chaotic laser output
${E_{\rm LD}^{(1)}(t)e^{i\omega_{\rm LD} t}, E_{\rm LD}^{(2)}(t)e^{i\omega_{\rm LD} t}, \ldots,}$ ${E_{\rm LD}^{(M)}(t)e^{i\omega_{\rm LD} t}}$,
when combined with the optical frequency comb $E_{\mathrm{PM}}(t)$ and heterodyned as described in Eq.~(S5), generates an independent WCBES, denoted by
${I_{\rm w}^{(1)}(t), I_{\rm w}^{(2)}(t), \ldots, I_{\rm w}^{(M)}(t)}$.

The statistical independence of these parallel WCBES units is evaluated using the cross-correlation matrix across channels, as shown in Supplementary Fig.~5a.
For $M = 200$, all off-diagonal elements cluster tightly around zero, confirming negligible inter-channel correlations for parallel WCBES generation.
Supplementary Fig.~5b presents the corresponding entropy rates for 200 channels.
The entropy rate is estimated using the NIST SP~800-90B suite~\citep{NIST.SP.800-90B-2018}, as described in the Methods section of the main text.
For consistency with the experimental conditions, the temporal waveforms
${I_{\rm w}^{(1)}(t), I_{\rm w}^{(2)}(t), \ldots,}$ ${I_{\rm w}^{(M)}(t)}$
are resampled to 256~GS/s (using the SciPy \texttt{resample\_poly} function) and digitized to 10 bits prior to entropy evaluation.
The estimated entropy rate per channel, $H_{\rm NIST}$, is approximately 1.86~Tb/s for 10-bit digitization, in agreement with the experimental result shown in Fig.~2b of the main text.
This implies that an aggregate entropy rate of approximately 372~Tb/s ($\approx 200$ channels~$\times$~1.86~Tb/s) can be achieved using 200 independent channels, as shown in Supplementary Fig.~5b.

The inset of Supplementary Fig.~5b shows the median value of $H_{\rm NIST}$ for $I_{\rm w}^{(1)}(t)$ to $I_{\rm w}^{(200)}(t)$ (red symbols) as a function of the number of digitized bits, together with the experimental results (black symbols) presented in Fig.~2b of the main text.
The simulated and experimental values of $H_{\rm NIST}$ are in close agreement, further supporting the accuracy of the simulations.
These results underscore the strong potential of the proposed architecture for ultrafast and massively scalable random bit generation.

\end{document}